**Metabolic alterations caused by smoking: the use of 1H-NMR in blood plasma analysis to unravel underlying mechanisms of lung cancer's leading risk factor.**

Goossens J[1], Gieghase J[1], Elien Derveaux[1], Peter Adriaensens[2]

[1]Department of Medicine and Life Sciences, Hasselt University, Hasselt, Belgium
[2]Applied and Analytical Chemistry, Institute for Materials Research, Hasselt University, Hasselt, Belgium

*Running Title: Performing metabolomics by 1H-NMR blood analysis*

Correspondence should be addressed to:
P. Adriaensens, Tel: +32 (11) 26 839 6, Email: peter.adriaensens@uhasselt.be

**Keywords: 1H-NMR, metabolomics, smoking, lung cancer, data reduction**

**ABSTRACT**

*Nowadays, the screening methods for the early detection of lung cancer struggle with several limitations such as many false positive results and low sensitivity. The detection of specific biomarkers is of high interest to complement these conventional screening methods. The objective of this study is to prove the power of 1H-NMR in metabolomics for the detection of smoking behavior, which is the leading risk factor for lung cancer, and as such gain more insights in the metabolic alterations that are caused by smoking. In this research, 1H-NMR spectra of human blood plasma samples were divided in 110 integration regions, from which the integration values were used to train an OPLS-DA classification model that underwent further data reduction. Results show that a classification model could discriminate between individuals based on their smoking status with a sensitivity of 96% and a specificity of 94%. This study also demonstrates that by performing a pathway-specific variable reduction of almost 50%, the sensitivity and specificity of the model almost remains the same. To conclude, 1H-NMR analyses show that the blood's metabolic profile of a smoker is altered compared to that of a non-smoker. Also, pathway-specific variable reduction shows great potential to perform overall data reduction. This workflow could be interesting to apply in the identification of lung cancer, to potentially detect specific biomarkers.*

**Introduction**

Lung cancer is the most prevalent cancer type worldwide, with more than 1.3 million cases each year and only a 5-year survival rate of around 15%.[1,2]

Lung cancer screening using Low-Dose Computed Tomography (LDCT) in high-risk individuals reduced lung cancer deaths by more than 20% when compared with those screened by chest radiography.[3] One of the main problems with the way that screening of lung cancer is currently done, is that patients are usually diagnosed at an advanced stage and thus limited curative treatment options are left.

Therefore, early detection could improve lung cancer survival rates.[4] Numerous lung cancer detection methods, such as Computed Tomography (CT), Chest Radiograph (CRG), Magnetic Resonance Imaging (MRI), Positron Emission Tomography (PET), as well as biopsy, have been used for screening purposes. However, these methods have limited screening capability, due to their low sensitivity, high cost, and physical/chemical damage.[5] Because of these limitations, other screening methods are being investigated.[6] To complement the current screening method, advances of high-throughput technology have shown potential in the detection of lung cancer biomarkers for early lung cancer detection.[6,7] One way this can be done is by determining an individual's metabolite profile.

It has been well known that the metabolism of cancer cells differs from that of healthy cells.[8] These alterations fulfill their different metabolic needs; e.g. to perform excessive proliferation. The metabolic changes are caused by genetic mutation that drive tumorigenesis. One of these key differences was described by Warburg, hence named the 'Warburg effect'. This effect describes the metabolic rewiring for the excessive conversion of glucose to pyruvate for enhanced production of lactate instead of entering the tricarboxylic acid (TCA) cycle.[9] Even in normoxic conditions, the cells rely on anaerobic energy production through glycolysis, hereby avoiding the slow oxidative phosphorylation by the TCA cycle. This reprogramming includes the overexpression of glucose transporters and glycolytic enzymes, high-speed adenosine triphosphate (ATP) production by branching off the glycolysis, and accumulation of lactate which drives tumor progression.[9]

Another major metabolic alteration in cancer cells is the increased demand for the amino acid glutamine. This increased demand for glutamine is attributed to its carbon backbone and its role in the replenishing of TCA cycle intermediates, called anaplerosis.[10] Besides, other studies are now emphasizing the importance of glutamine in cancer as proteogenic building block, a nitrogen donor and carrier, an exchanger for import of other amino acids, and a signaling molecule.[11] All in all, it's been known that the synthesis of glutamine is highly being altered in cancerous cells. By understanding these metabolic changes, these hallmarks can potentially be used in the detection of cancer by metabolomics.[10]

The reprogramming by the cancer cells fulfill their different metabolic needs. The combination of all these effects can be seen in the concentrations of the corresponding metabolites. The altering of these concentration differences can be an interesting measure for the detection of cancer.[12]

Metabolomics is the study of all small, low-weight molecules originating from a cell's metabolism. These cellular processes can be specified with a unique metabolic fingerprint.[13] This powerful discipline allows us to determine all small metabolites and their concentration present in a biological sample.[14]

The two main techniques used in metabolomics are Mass Spectrometry (MS) and Nuclear Magnetic Resonance (NMR). Proton- ($^1$H)-NMR is mostly studied, as all hydrogen-containing molecules, which all metabolites are, can together result in a highly informative NMR spectrum based on the proton nucleus. Compared to MS,[1]H-NMR has the advantages of being nondestructive, in need of minimal sample preparation, cost-effective and reproducible with a rapid high-throughput data acquirement. One NMR measurement of only a few minutes, can acquire information about all metabolites present in the sample.[15] On the other hand, MS is more sensitive, but requires the metabolites to be separated before detection, therefore often combined with Liquid Chromatography (LC-MS).[16]

The general principle of [1]H-NMR is to change the core spin rotation of a hydrogen atom by application of an external magnetic field. As these electrically charged hydrogen atoms spin around their axis, a magnetic moment (μ) is created for each hydrogen atom *(Figure 1A)*.[17,18] Additionally, a torque is created by the magnetic moments (also called precession). This Larmor precession, with a specific Larmor frequency, is the movement of the nuclei around the direction of the magnetic field *(Figure 1A)*.[19]

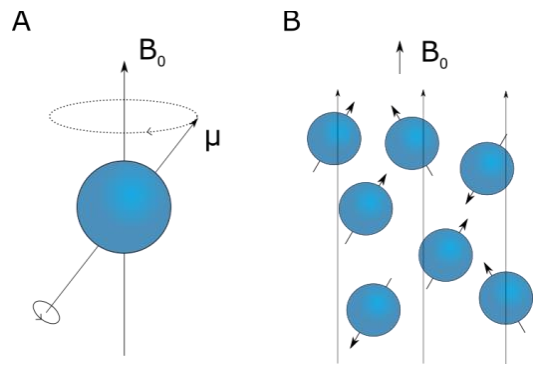

*Figure 1: (A) Magnetic moment (μ) created by the spin of the hydrogen atom, which precesses around the axis of the magnetic field ($B_0$). (B) These magnetic moments will align when an external magnetic field is applied.*

In a random orientation, the overall net magnetic moment ($M_0$) is zero, but when an external magnetic field ($B_0$) is applied, all the magnetic moments (μ) align perpendicular towards the direction of this field. In this conformation, they can possess two distinctive states: α-state (parallel to $B_0$) or β-state (anti-parallel to $B_0$) *(Figure 1B)*. As the protons in α-state are at a lower energy level, more protons are in the α- than β-state. Therefore, the net magnetization results in a vector aligning with the z-axis *(Figure 2)*.[18,19]

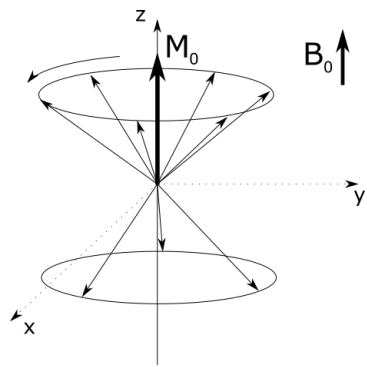

*Figure 2: Net magnetization ($M_0$) aligning with the z-axis, produced by parallel and anti-parallel oriented magnetic moments ($\mu$) from hydrogen atoms. As more parallel than anti-parallel magnetic moments are present, the net magnetization results in a positive value in the z-axis.*

After applying a well-defined radio frequency wave (known as a 90° pulse), the magnetic moments are forced into phase coherence. This means that all the energy is being absorbed by the protons, and all their magnetic moments align with the y-axis (90° pulse) *(Figure 3)*. Subsequently, this 90° pulse is accurately set to acquire equal populations of protons with parallel and anti-parallel magnetic moments. This results in the removal of the z-component of the net magnetization ($M_0$), which leaves it completely in the x-y plane.[18,19]

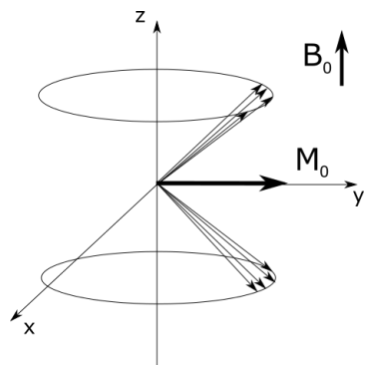

*Figure 3: Net magnetization ($M_0$) aligned with the y-axis as result of a 90° pulse.*

After this pulse, $M_0$ is maximal in the y-axis but reduces over time due to relaxation and fanning out processes. Due to the flip of anti-parallel to parallel state protons, the protons relax to a lower energy state, resulting in the decrease of the y-component of $M_0$ while increasing its z-component again (Relaxation/$T_1$ recovery) *(Figure*

*4A)*. All the magnetic moment in the x-y plane precess at slightly different frequencies.

After the pulse, the magnetic moments will fan out at different frequencies and lose their phase coherence, resulting in the decrease of the y-component of $M_0$ (Fanning out/$T_2$ decay) *(Figure 4B)*.[18,19]

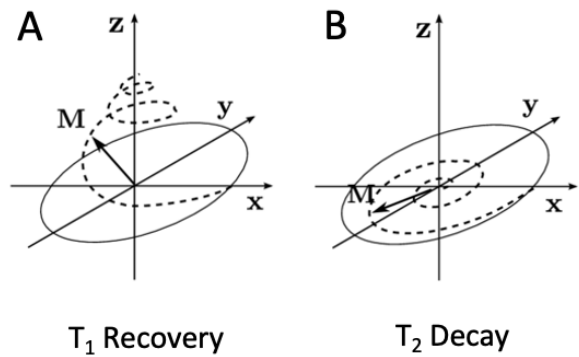

*Figure 4: (A) $T_1$ recovery: increase of the net magnetization vector ($M_0$) in the z-component after the 90° pulse due to the flip of anti-parallel to parallel oriented magnetic moments. (B) $T_2$ decay: decrease of the net magnetization vector ($M_0$) in the y-component after the 90° pulse due to the fanning out of the magnetic moments at different frequencies.[20]*

The overall signal decrease of the y-component is measured by a receiver coil and results in a Free Induction Decay (FID). This decay is linked to the emission of specific electromagnetic radiation emitted by each proton, with a magnitude dependent on the chemical environment of each proton. An additional Fourier transformation is applied to scale time to frequency. After this mathematical transformation, a typical $_1$H-NMR spectrum is obtained, where the magnitude is plotted in function of chemical shift (and additionally a magnetic field-independent scale, expressed in Parts Per Million (ppm)) *(Figure 5)*.[18,19,21]

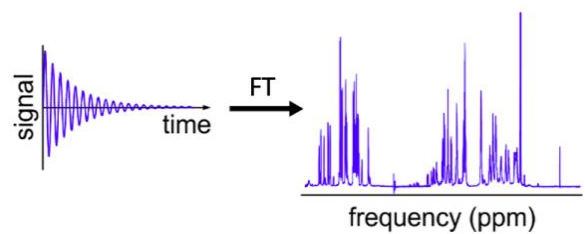

*Figure 5: Typical $_1$H-NMR spectrum of blood plasma, which is the result of the Fourier-transformated FID.[21]*

Since human blood consists of more than metabolites alone, the 1H-NMR spectrum will also contain an enormous number of signals derived from macromolecules (e.g. lipids, polysaccharides…). Therefore, a specific pulse sequence is needed to suppress all signals from macromolecules and focus on the signals of metabolites alone. Different compounds can be filtered out based on a difference in relaxation time. Macromolecules have a faster relaxation time (T2) than the slower relaxing metabolites *(Figure 6)*.[18]

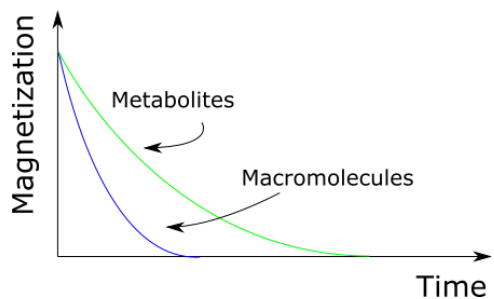

*Figure 6: Difference in relaxation time (T2) for macromolecules and metabolites.*

To reduce the signal of these faster relaxing magnetizations, a Carr-Purcell-Meiboom-Gill (CPMG) pulse sequence is being administered. This CPMG pulse starts with a 90° pulse, followed by a loop of an even amount of 180° pulses after which the acquisition can occur. As mentioned earlier, after the 90° pulse, all the magnetic moments are in phase, but these will start to fan out in the x-y plane over time. When the 180° pulse is given, the magnetic moments are being flipped around y-z plane and start to get into phase again. Thus, the magnetic moments that precess faster than the Larmor frequency will catch up to the rotating coordinate system (rotating with the Larmor frequency). The rotating coordinate system will now in its place catch up to the slower precessing magnetic moments *(Figure 7)*. Additionally, the CPMG pulse takes care of possible inhomogenities from imperfect 180° pulses. This is also the reason why the pulses are administerd in even amounts.[18]

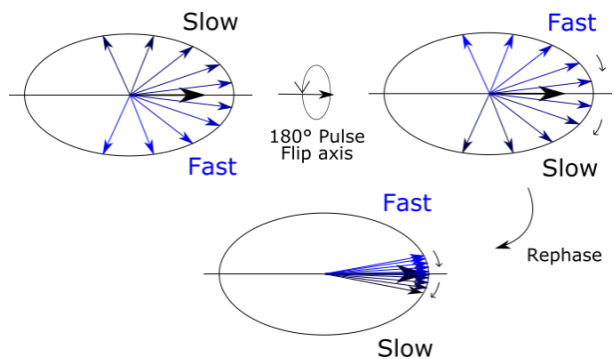

*Figure 7: CPMG pulse: Catching up of slower magnetic moments after the axis get flipped in a 180° pulse.*

The research group of prof. Adriaensens used 1H-NMR as an analytical tool to perform metabolomics on human blood plasma for the detection of lung cancer.[22] A 400 MHz NMR spectrometer was used to determine the human plasma metabolic profiles of lung cancer patients (n=233) and healthy controls (n=226). Subsequently, the obtained spectra were divided into 110 integration regions and on their turn, these integration values were used to train a statistical Orthogonal Partial Least Squares Discriminant Analysis (OPLS-DA) model. The specific limits of these integration regions were based on an earlier study by Louis et al. Here, human blood plasma was spiked with 32 different metabolites and these metabolite signals were assigned to different integration regions.[23] The resulting OPLS-DA model was afterwards validated by an independent cohort consisting of 98 lung cancer patients and 89 controls, and a sensitivity of 78% (meaning that 78% of the lung cancer patients were correctly classified) and specificity of 92% (meaning that 92% of the healthy controls were correctly classified) could be achieved.

This study was performed using a medium-field (400 MHz) spectrometer. As reviewed by Derveaux et al., when using an NMR spectrometer with a higher magnetic field strength (800-900 MHz), an improved resolution as well as a higher signal to noise ratio can be obtained. Because of these improvements, the integration regions can be determined more accurately when performing spiking experiments, resulting in less signal overlap and therefore in a larger number of integration regions that can be assigned to a single metabolite. The main drawback of these high-field spectrometers is the high cost, and the demand for an isolated building for its housing. Therefore, with the focus on clinical metabolomics, medium-field (400-600 MHz) spectrometers will probably be used in the future.[15]

The goal of this research is to investigate the power of $_1$H-NMR in metabolomics for the production of OPLS-DA classification models. Besides cancer causing a difference in the blood's metabolism, smoking does too.[24] Therefore, OPLS-DA models will be trained with the focus on differentiating individuals on the basis of their smoking status as lung cancer leading risk factor (1). To reduce external influences, only the control group of the lung cancer study will be used to train this model. At the time of sample collection, the smoking status (active, never, or stopped for at least six months) was included in the database. Subsequently, data reduction will be carried out on this model in a forward and backward approach (2). The aim is to determine if all 110 variables are essential to produce a good classification model and which ones are crucial for obtaining a good classification. Subsequently, the most contributing variables can be linked to their corresponding metabolites. At last, a search will take place for the identification of which pathways in the blood's metabolism are highly altered by an active smoking lifestyle. Therefore, the contribution of different pathways to the model will be studied (3).

**Materials and Methods**

*Subjects* – The same control subjects, as used in Louis et al., were used in this project. [22] Controls (n=347), with noncancer diseases, were included at Ziekenhuis Oost-Limburg (Genk, Belgium) between March 2012 and June 2014. None had malignant diseases, which was double checked on statistical misclassified controls. The control population was used for training and validation of the classification models for different smoking statuses, which were categorized in three groups (active, never, and stopped) *(Figure 8)*. Subjects were accounted as stopped smokers if they stopped more than 6 months before inclusion.

To reduce external factors, different exclusion criteria were introduced. Patients with following criteria were excluded from the study: (1) not fasted for at least 6 hours, (2) fasting blood glucose of 200 mg/dL or higher, (3) medication intake on the morning of blood sampling, (4) treatment or history of cancer in the past five years.

*Blood sampling and processing* – Fasting venous blood samples were collected and stored at 4° C within 5 minutes after collection. After 8 hours of blood collection, the samples were centrifuged at 1600 *g* for 15 minutes, and 500 µl of plasma was transferred into sterile cryovials for storage at – 80° C.

*Sample preparation* – Plasma aliquots were thawed, followed by centrifugation at 13000 rpm for 4 minutes at 4° C. 200 µl of the supernatants was resuspended in 600 µl $D_2O$ containing 0.3 µg/ml trimethylsilyl-2,2,3,3-tetradeutero-propionic acid (TSP) as a chemical shift reference (0.000 ppm). This solution was afterwards transferred to a 5 mm NMR tube.

*NMR analysis* – $_1$H-NMR spectra were recorded after a stabilization period of 7 minutes at 21.2° C on a 400 MHz Jeol spectrometer. Spectra recordings were started by an initial preparation delay of 0.5 s and 3 s presaturation for water suppression, 6000 Hz spectral width, an acquisition time of 1.1 s, 13K data points, and 96 scans. A CMPG pulse was used to suppress the macromolecules present in the plasma samples (total spin echo time of 32 ms and interpulse delay of 0.1 ms). Before Fourier transformation, the FID was zero-filled to 65 K points and multiplied by a 0.7 Hz exponential line-broadening function.

*Spectral processing* – Based on the chemical shift of the spiked metabolites, the $_1$H-NMR spectra were divided into 110 integration regions, excluding the water region (4.7-5.1 ppm) and TSP (-0.3-0.3 ppm). To check whether no positional changes occurs in the spectra due to mistakes in the NMR settings, a reference sample was measured before analyzing an unknown sample. This reference sample was then laid on top of the reference sample from the day before. If those two didn't match perfectly, the settings (mostly shimming) were optimized until they did. Especially the region that contain well-resolved, non-overlapping signals were examined (3.4-4 ppm). After recording the spectra, the integration values were normalized relatively to the total integrated areas, resulting in 110 normalized integration values, being the variables used for multivariate statistical analysis. This complete protocol is based on a previously reported method by Louis et al.[26]

*Statistical analysis* – Multivariate statistical analysis was performed using SIMCA-P+ (Version 14, Umetrics, Malmö, Sweden). After centering and Pareto scaling of all variables, Orthogonal Partial Least Squares Discriminant Analysis (OPLS-DA) was used to train a classification model. This classification model was validated on an independent dataset. The contribution of variables was analyzed by the use of S-plots, S-lines, and Variable importance of Projection (VIP) plots. Data reduction was based on the most contributing variables (highest VIP values). Each produced model was evaluated on model parameters indicating their explained variation within and between groups (R2X and R2Y), and predictive capacity (Q2), sensitivity, and specificity). A R2X value close to 1 means that the variability in the group is low, and the differences between the two groups are responsible for the separation of them. A R2Y value close to 1 means that the two groups are separated on the basis of the produced model. A Q2 value close to 0.5 means, in NMR-metabolomics research, that the model has an excellent predictive accuracy. Respectively, sensitivity and specificity indicate the percentage of correctly classified subjects as true positives and negatives.

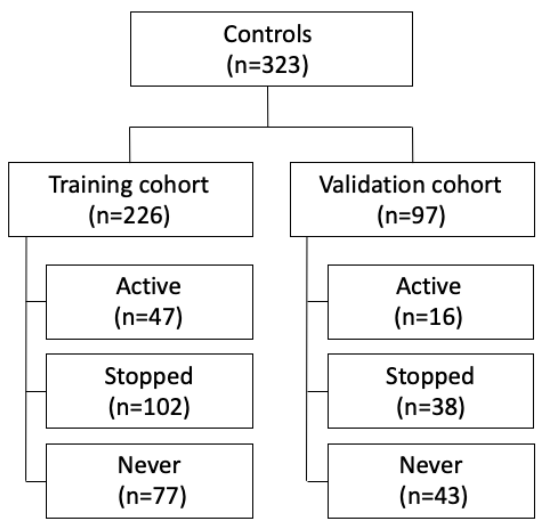

*Figure 8: CONSORT diagram for the study of smoking habit.*

## Results and Discussion

*Differentiation by smoking behavior* – The blood's metabolism changes by an individual's smoking behavior.[24] These metabolic differences in plasma were examined using $_1$H-NMR spectroscopy whereof the integration data of 110 variables was used to construct statistical classification models. With data from 226 healthy controls, three different OPLS-DA classification models were build based on two different smoking statuses each: Active vs. Never / Active vs. Stopped / Stopped vs. Never, and afterwards evaluated by an independent validation dataset of 97 subjects (*Table 1*). By evaluating these models, a few aspects can be deduced. First of all, as expected, based on the input of metabolic data, it is easier to differentiate between active smokers and patients who had never smoked than patients who quitted smoking. The training model 'Active vs. Never' shows a very good classification between the two groups with a sensitivity of 95.74% and specificity of 93.51%. Furthermore, the validation model 'Active vs. Never' had the highest predictive values, with a sensitivity of 62,5 % and a specificity of 88.37 %. However, it should be noted that these values might increase when more subjects are added to the validation cohort. The classification models of 'Active vs. Stopped' or 'Stopped vs. Never' had less predictive values. It makes sense that a classification model can easier differentiate between two more extreme cases, which explains why 'Active vs. Never' was the best classification model. Despite being less accurate, the model of 'Active vs. Stopped' can somewhat differentiate between the two groups. The predicative values of the model of 'Never vs. Stopped' are too low to correctly differentiate between the two groups. Additionally, as 'Active vs. Stopped' had the second-best predictive values, which is still quite low, it can be deduced that a stopped smoker's metabolism can better be resembled by that of a non-smoker than a smoker.

*Table 1: Characteristics of the classification models for smoking behavior.*

|  | Var | LV (P + O) | R2X(Cum) | R2Y(Cum) | Q2(Cum) | Sens (%) | Spec (%) |
|---|---|---|---|---|---|---|---|
| *Training data* | | | | | | | |
| Active vs. Never | 110 | 8 (1 + 7) | 0.919 | 0.687 | 0.322 | 95.74 | 93.51 |
| Active vs. Stopped | 110 | 6 (1 + 5) | 0.890 | 0.512 | 0.263 | 93.14 | 78.72 |
| Never vs. Stopped | 110 | 3 (1 + 2) | 0.768 | 0.122 | 0.00792 | 78.43 | 51.95 |
| *Validation data* | | | | | | | |
| Active vs. Never | 110 | - | - | - | - | 62.50 | 88.37 |
| Active vs. Stopped | 110 | - | - | - | - | 81.58 | 56.25 |
| Never vs. Stopped | 110 | - | - | - | - | 73.68 | 25.58 |

Var, number of variables; LV, latent variable; O, number of orthogonal components; P, number of predictive components; R2X(Cum), total explained variation in X; R2Y(Cum), total explained variation in Y; Q2(Cum), predicted variation; Sens, sensitivity, Spec, specificity.

Since the best values were obtained from the model that discriminates between the smoking behavior 'Active vs. Never', this will be used in following experiments. Importantly, possible confounders of the model 'Active vs Stopped' were investigated in the validation data. The clinical parameters of BMI, age and gender are visualized in OPLS-DA plots, from which the distribution can be examined. These three parameters show an even distribution in the overall plot, which indicated that these are no confounders (*SI Figure 1*).

*NMR spectra analysis* – As earlier mentioned, the $_1$H-NMR spectra were processed by dividing the spectra in 110 integration regions, and the integration values were automatically determined by the software. For illustrating purposes, two differentiating integration regions were visually examined. An active, stopped, and non-smoker were chosen on the basis of an OPLS-DA score plot, where three clear cases were selected (*Figure 9*). The NMR spectra of these patients were placed on top of each other, with special attention to two integration regions (1.3740 – 1.3450 ppm and 2.1230 – 1.9720 ppm) where two interesting phenomena can be seen (*Figure 10*). In the first example (*Figure 10A*), it can be seen that the signal at 2.11-2.12 shows a clearly sharp signal for the active and stopped smoker, which is not the case for the non-smoker. It's clear that it is difficult to visually examine these integration regions. Not all integration differences can easily be spotted, let alone be quantified. Subsequently, there is a small variation between different recorded spectra, demonstrated as another difficulty in the second example (*Figure 10B*). One could expect that here the differences between the spectra could be appointed to the difference in smoking behavior. Actually, this is a example of a baseline shift. It's not likely that there is a difference in peak height, but the baseline, the level where the peak starts to rise, is shifted. This phenomenon can also be appointed for the differences between the signals from other peaks in the first example (e.g. 1.98-2.04 in *Figure 10A*). The Jeol software, which determines the integration values, can automatically perform baseline shift corrections at specific regions. It has to be clear that it's difficult to visually determine the difference and a machine learning-based program is needed to gain reliable results.

*Data reduction* – The models that were built so far, consisted of all 110 variables that were obtained from the $_1$H-NMR spectra. But one has to be aware of the fact that not all 110 variables have an equal contribution to the classification model. Some metabolites (and therefore variables) play a more important role. By investigating which variables contribute the most to the model, one can detect which metabolites (or even biochemical pathways) are influenced by smoking. On the other hand, data reduction is also important from a statistical standpoint. When input is large and consist of enough data, machine-learning software can almost always construct a classification model. However, such a model will not hold for a new, independent data cohort. Therefore, the goal will be to prove the power of this technique while reducing the number of variables. Data reduction of the model on smoking behavior 'Active vs. Never' will be carried out by a backward and forward approach.

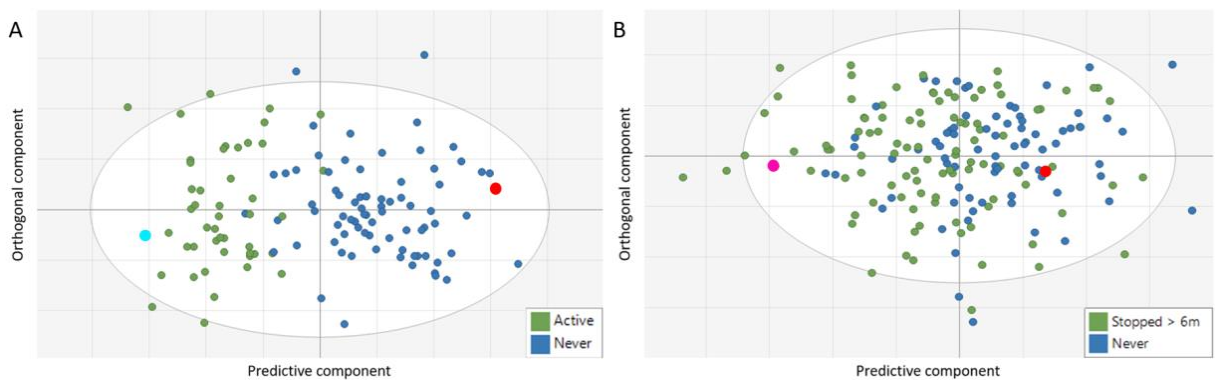

*Figure 9: OPLS-DA score plot for models on smoking behavior (A: Active vs. Never, B: Stopped vs. Never). Three clear cases for each smoking behavior were selected (Active: light blue, Never: red, Stopped: Pink).*

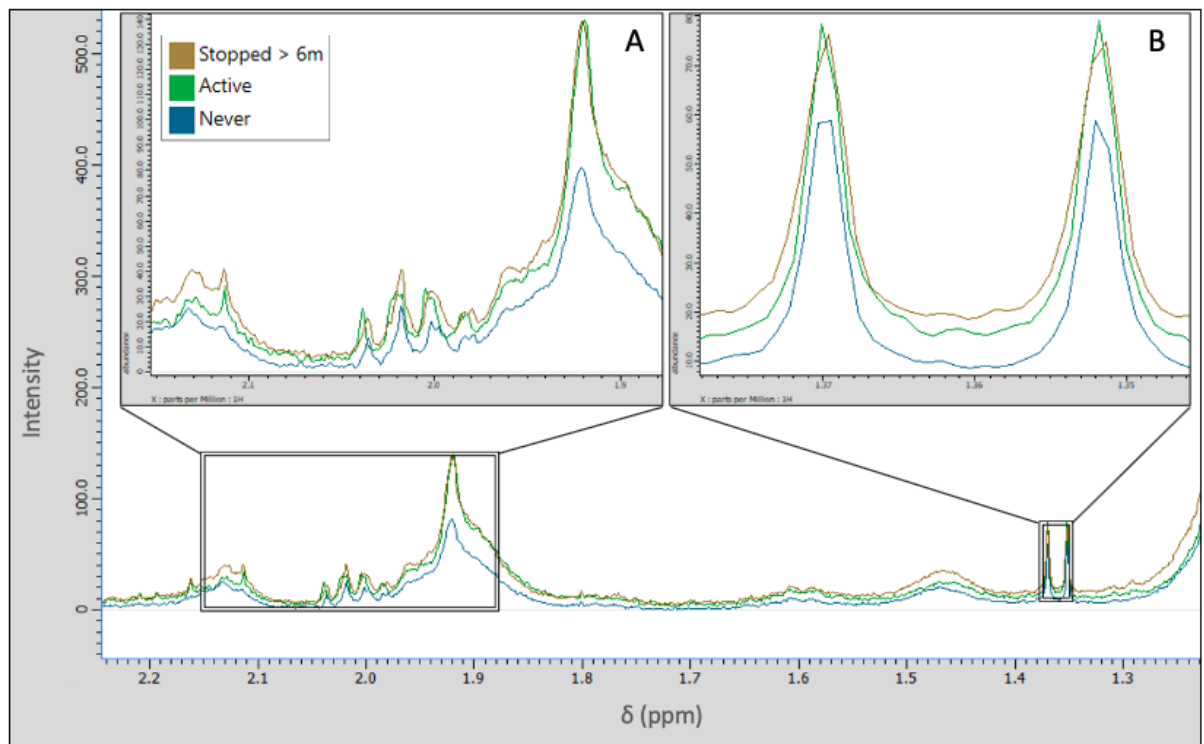

*Figure 10: NMR spectra analysis for an active (green), stopped (brown), and non- smoker (blue) with special attention to two integration regions ((A) 2.1230 – 1.9720 and (B) 1.3740 – 1.3450).*

*Backward data reduction* – To examine the contribution of the variables to the model, a VIP plot was used. The variables with the least contribution (lowest VIP value) were excluded from the new dataset. This dataset was then used to train a new classification model and the prediction values were determined. By excluding 10/110 variables, the predictive capacity on the training dataset didn't alter much, and the sensitivity and specificity on the validation data remained exactly the same. By excluding more and more variables (20, 25, 30), R2X, R2Y, and Q2 values slowly decreased (*Table 2*). In order to reduce the variables, it also could be interesting to exclude all metabolites belonging to a complete pathway that would experience no impact from a changing smoking behavior. The corresponding metabolites for the 30 least contribution variables were traced back, and matching pathways were determined. Here, it could be seen that these metabolites do not appertain to the same pathway, and therefore makes it impossible to exclude one (or more) specific pathway(s).

*Forward data reduction* – The forward approach was carried out based on the same method as the backwards approach. Firstly, classification models were trained based on the most contributing variables (10, 20, 25, 30; excluded based on VIP values). Afterwards, a search to find specific pathways responsible for the change in metabolism was performed. As expected, models trained with only a small number of variables show a lower predictive power, but this might give an indication of how important the first variables in the models actually are. In *Table 2*, the predictive values of the models are shown, from which a couple aspects can be deduced. Interestingly, the model with only 10 variables shows an R2X value of 0.995, meaning that the variability within the classification groups is extremely low. The variability increases within the group when adding more variables (except the models constructed with almost every variable). Additionally, the model trained with only 25 variables shows reasonable high predictive values compared to the models trained with 30 (and even 80) variables. Here, R2Y and Q2 are slightly higher, meaning that the model with 25 variables can better differentiate between the active and non-smokers than a model with 30 (or even 80) variables. It seems like it is very important to know which variables needs to be in- or excluded. It could be that, the metabolism is highly influenced by certain specific pathways, and the co-operation of these variables play an important role.

*Table 2: Characteristics of the classification models for backwards and forwards data reduction on the basis of variables with the least and most contribution to the model.*

| | Var | LV (P + O) | R2X(Cum) | R2Y(Cum) | Q2(Cum) | Sens (%) | Spec (%) |
|---|---|---|---|---|---|---|---|
| **Training data** | | | | | | | |
| Active vs. Never | 110 | 8 (1 + 7) | 0.919 | 0.687 | 0.322 | 95.74 | 93.51 |
| Active vs. Never | 100 | 8 (1 + 7) | 0.923 | 0.673 | 0.316 | 93.62 | 90.91 |
| Active vs. Never | 90 | 5 (1 + 4) | 0.884 | 0.472 | 0.232 | 72.34 | 88.31 |
| Active vs. Never | 85 | 5 (1 + 4) | 0.887 | 0.459 | 0.226 | 68.09 | 89.61 |
| Active vs. Never | 80 | 4 (1 + 3) | 0.857 | 0.387 | 0.207 | 61.70 | 81.82 |
| Active vs. Never | 30 | 5 (1 + 4) | 0.943 | 0.380 | 0.227 | 61.70 | 88.31 |
| Active vs. Never | 25 | 6 (1 + 5) | 0.967 | 0.422 | 0.259 | 74.47 | 88.31 |
| Active vs. Never | 20 | 6 (1 + 5) | 0.970 | 0.419 | 0.269 | 74.47 | 88.31 |
| Active vs. Never | 10 | 7 (1 + 6) | 0.995 | 0.394 | 0.262 | 76.60 | 87.01 |
| **Validation data** | | | | | | | |
| Active vs. Never | 110 | - | - | - | - | 62.50 | 88.37 |
| Active vs. Never | 100 | - | - | - | - | 62.50 | 88.37 |
| Active vs. Never | 90 | - | - | - | - | 62.50 | 86.05 |
| Active vs. Never | 85 | - | - | - | - | 62.50 | 86.05 |
| Active vs. Never | 80 | - | - | - | - | 56.25 | 81.04 |
| Active vs. Never | 30 | - | - | - | - | 50.00 | 83.72 |
| Active vs. Never | 25 | - | - | - | - | 62.50 | 81.40 |
| Active vs. Never | 20 | - | - | - | - | 56.25 | 83.72 |
| Active vs. Never | 10 | - | - | - | - | 43.75 | 76.74 |

Var, number of variables; LV, latent variable; O, number of orthogonal components; P, number of predictive components; R2X(Cum), total explained variation in X; R2Y(Cum), total explained variation in Y; Q2(Cum), predicted variation; Sens, sensitivity, Spec, specificity.

Therefore, in the search of important contributing pathways, like earlier demonstrated in the backward approach, the metabolites corresponding to the first 10 variables were traced back. These metabolites were then used to determine matching pathways where the following was found: The 10 variables corresponding to 16 different metabolites, matched in 6 different pathways (pathways were appointed to match if at least three metabolites were listed). The matching pathways are: 1) valine, leucine and isoleucine biosynthesis and 2) degradation, 3) cysteine and methionine metabolism, 4) alanine, aspartate and glutamine metabolism, 5) 2-oxocarboxylic acid metabolism and 6) cyanoamino acid metabolism. Next, all the metabolites for these previous mentioned pathways were listed, and all the variables corresponding to these metabolites were traced back. Because the 2-oxocarboxylic acid and cyanoamino acid metabolism pathways are enormous, these were excluded from this experiment. As all variables corresponding to these pathways would be studied, almost all 110 variables would be investigated, which is not the goal of this experiment. Therefore, only the first four pathways will be studied, resulting in a model of 58 variables, which is still a variable reduction of almost 50%. At last, a model was trained with these 58 variables, and compared with the original model. Surprisingly, with this model of only 58 variables, the sensitivity and specificity on the validation data decreased only a little compared to the original model of 110 variables. On the other hand, the power of the training model (R2X, R2Y, and Q2) had decreased severely (*Table 3*). (*Please refer to SI Table 1 for the metabolites, and the pathways wherein they specifically appear, which were needed to build the models*)

In order to know which pathway contributed the most, different models were built for each and every pathway itself (*Table 3*). The data showed, that only models build on the pathways 'cysteine and methionine metabolism' and 'alanine, aspartate and glutamine metabolism' could slightly differentiate between active or non-smokers. Although the power of these models is low, the R2Y and Q2 values of the models on 'valine, leucine and isoleucine biosynthesis and degradation' are extremely low, meaning these models can't differentiate by any means between active or non-smokers. Despite having a lower power (R2X, R2Y) and predictive accuracy (Q2), the sensitivity and specificity of the 'alanine, aspartate and glutamine metabolism'-model was even higher than that of all pathways together. To sum up, it can be expected that the pathways of 'cysteine and methionine metabolism' and 'alanine, aspartate and glutamine metabolism' are highly influenced by an individual's smoking behavior.

*Table 3: Characteristics of the classification models for forward data reduction on the basis of pathway specific contribution variables.*

| | Var | LV (P + O) | R2X(Cum) | R2Y(Cum) | Q2(Cum) | Sens (%) | Spec (%) |
|---|---|---|---|---|---|---|---|
| **Training data** | | | | | | | |
| Original | 110 | 8 (1 + 7) | 0.919 | 0.687 | 0.322 | 95.74 | 93.51 |
| Reduced (VIP) | 58 | 4 (1 + 3) | 0.875 | 0.357 | 0.204 | 61.70 | 84.42 |
| Reduced (Path.) | 58 | 4 (1 + 3) | 0.873 | 0.360 | 0.189 | 65.96 | 84.42 |
| Pathway 1 | 33 | 2 (1 + 1) | 0.832 | 0.172 | 0.0844 | 40.43 | 81.82 |
| Pathway 2 | 29 | 1 (1 + 0) | 0.749 | 0.0838 | 0.0645 | 36.17 | 88.31 |
| Pathway 3 | 22 | 3 (1 + 2) | 0.835 | 0.296 | 0.140 | 59.57 | 85.71 |
| Pathway 4 | 27 | 3 (1 + 2) | 0.820 | 0.328 | 0.196 | 63.83 | 84.42 |
| **Validation data** | | | | | | | |
| Original | 110 | - | - | - | - | 62.50 | 88.37 |
| Reduced (VIP) | 58 | - | - | - | - | 56.25 | 79.07 |
| Reduced (Path.) | 58 | - | - | - | - | 62.50 | 81.40 |
| Pathway 1 | 33 | - | - | - | - | 50.00 | 74.42 |
| Pathway 2 | 29 | - | - | - | - | 31.25 | 88.37 |
| Pathway 3 | 22 | - | - | - | - | 43.75 | 83.72 |
| Pathway 4 | 27 | - | - | - | - | 68.75 | 81.40 |

Var, number of variables; LV, latent variable; O, number of orthogonal components; P, number of predictive components; R2X(Cum), total explained variation in X; R2Y(Cum), total explained variation in Y; Q2(Cum), predicted variation; Sens, sensitivity; Spec, specificity; Path., Pathways: 1, valine, leucine and isoleucine biosynthesis; 2, valine, leucine and isoleucine degradation; 3, cysteine and methionine metabolism; 4, alanine, aspartate and glutamine metabolism.

*Predictive power of pathway specific modelling* - In order to assign the predictive power to these specific pathways, this pathway-based model of 58 variables was compared to a reference model. This reference model contained the 58 most contributing variables of the original model, irrespectively of any pathway. The 58 most contributing variables were determined by their VIP values. Although, the training models were comparable, the pathway-based model had a higher sensitivity and specificity on the validation dataset. Interestingly, 21 of the 58 variables were different, meaning that the pathway-specific variables are as important as the statistically assigned most contributing variables.

It's been well known that smoking influences the blood metabolism. Different MS-based studies could already demonstrate that a smoker's blood metabolism is highly characteristic, and different from a non-smoker.[27–29] Gu et al. managed to assign the difference in 25 metabolite concentrations to smoking, but no further investigation took place which specific pathways were altered.[24] On the other hand, Xu et al. could assign the difference in 21 metabolite concentrations to smoking, which showed enrichment in a set of amino acid and lipid metabolism pathways (ether lipid, glycerol-phospholipid, arginine and proline metabolism).[30] Also, their results showed that 19 out of the 21 metabolite differences were found to be reversible in former smokers, which indicated the changes in human serum metabolites are reversible after smoking cessation, which is in accordance to our results.[30]

**Future prospects**

In further studies, it could be interesting to investigate how the smoking-specific pathways are altered, and which metabolites are potential key players. Hereby, more insight will be gained by the alterations in human blood metabolism caused by smoking. Therefore, more spiking experiments with an additional set of important metabolites could be performed, in order to examine the pathways more thoroughly and specifically. As not all integration regions of all existing metabolites are known, there still is a loss of information which otherwise could potentially improve the current setup.

Additionally, if a 600 MHz spectrometer would be used, a more detailed $1H$-NMR spectrum would be obtained, and the integration regions could better be determined since resolution will be increased and thus more integration regions corresponding to a single metabolite might be obtained. Such a change in the dataset of variables could improve the predicative power of the OPLS-DA classification models.

This study focused on the discrimination of individuals based on their smoking behavior, which is known to be an enormous risk factor of lung cancer development. The next step would be to determine the pathways that are specifically altered by lung cancer and investigate the link between the altered pathways of smoking and the altered pathways of lung cancer development. Potentially, the effects of smoking on the blood's metabolism could be linked to the development of lung cancer. It would be ideal to find specific pathways that indicate the early onset of lung cancer development. If specific biomarkers would be discovered, $_1$H-NMR could potentially be used as a complementary screening method to improve the screening effectiveness for lung cancer of today.

## Conclusion

In conclusion, this study validates the power of $_1$H-NMR in metabolic-based phenotyping of human blood plasma as a tool to discriminate individuals on the basis of their smoking behavior.

Application of multivariate statistics resulted in an OPLS-DA model that allows discrimination between smokers and non-smoker with a sensitivity of 96% and a specificity of 94%. Therefore, it could be deduced that a smoker's blood metabolism differs from that of a non-smoker. Additionally, a quitted smoker's metabolism shows more similarities to a non-smoker's than that of an active smoker. This might mean that the human body is able to recover from the metabolic alterations done by smoking on the blood's metabolism.

Data reduction was performed in a backward and a forward manner. The backwards approach showed clearly that not all variables were essential to construct a good classification model. On the other hand, the forward approach showed a greater potential in the assignment of specific pathways that were altered by active smoking.

It could be deduced that some pathways highly contribute to the differences in the blood's metabolism. Especially the pathways of 'cysteine and methionine metabolism' and 'alanine, aspartate and glutamine metabolism' are thought of playing an important role. Even with a variable reduction of almost 50%, the sensitivity and specificity remained almost the same.

In the future, it could be interesting to apply this method for the detection of specific altered biochemical pathways in the early development of lung cancer. Potentially, the effects of smoking on the blood's metabolism could be linked to the early development of lung cancer.

## Acknowledgements

The authors are grateful for the technical and theoretical support provided by the university of Hasselt (UHasselt). Special thanks to the personnel of the Department of Medicine and Life Sciences and the research group of Applied and Analytical Chemistry for the personal advice and support in performing this research.

## References

1. Field, J. K. Lung cancer risk models come of age. *Cancer Prev. Res. (Phila).* **1**, 226–8 (2008).
2. Ferlay, J. *et al.* Cancer incidence and mortality worldwide: Sources, methods and major patterns in GLOBOCAN 2012. *Int. J. Cancer* **136**, E359–E386 (2015).
3. Usman Ali, M. *et al.* Screening for lung cancer: A systematic review and meta-analysis. *Preventive Medicine* vol. 89 301–314 (2016).
4. Inage, T., Nakajima, T., Yoshino, I. & Yasufuku, K. Early Lung Cancer Detection. *Clinics in Chest Medicine* vol. 39 45–55 (2018).
5. Wang, L. Screening and Biosensor-Based Approaches for Lung Cancer Detection. *Sensors* **17**, 2420 (2017).
6. Hasan, N., Kumar, R. & Kavuru, M. S. Lung cancer screening beyond low-dose computed tomography: The role of novel biomarkers. *Lung* vol. 192 639–648 (2014).
7. Yang, G. *et al.* Recent advances in biosensor for detection of lung cancer biomarkers. *Biosens. Bioelectron.* **141**, 111416 (2019).
8. Cantor, J. R. & Sabatini, D. M. Cancer cell metabolism: One hallmark, many faces. *Cancer Discovery* vol. 2 881–898 (2012).
9. Vaupel, P., Schmidberger, H. & Mayer, A. The Warburg effect: essential part of metabolic reprogramming and central contributor to cancer progression. *International Journal of Radiation Biology* vol. 95 912–919 (2019).
10. Sciacovelli, M., Gaude, E., Hilvo, M. & Frezza, C. The metabolic alterations of cancer cells. in *Methods in Enzymology* vol. 542 1–23 (Academic Press Inc., 2014).
11. Bott, A. J., Maimouni, S. & Zong, W. X. The pleiotropic effects of glutamine metabolism in cancer. *Cancers* vol. 11 (2019).
12. Emwas, A. H. *et al.* Nmr spectroscopy for metabolomics research. *Metabolites* vol. 9 (2019).
13. Growing Pains for Metabolomics | The Scientist Magazine®. https://www.the-scientist.com/technology/growing-pains-for-metabolomics-48835.
14. Hollywood, K., Brison, D. R. & Goodacre, R. Metabolomics: Current technologies and future trends. *Proteomics* vol. 6 4716–4723 (2006).
15. Derveaux, E. *et al.* Diagnosis of Lung Cancer: What Metabolomics Can Contribute. *Lung Cancer - Strateg. Diagnosis Treat.* (2018) doi:10.5772/intechopen.79258.
16. Nicholson, J. K. & Lindon, J. C. Systems biology: Metabonomics. *Nature* vol. 455 1054–1056 (2008).

17. A more precise measurement of the proton's magnetic moment - Resonance Science Foundation. https://resonance.is/precise-measurement-protons-magnetic-moment/.

18. Shoulders, B. *High-Resolution NMR Techniques in Organic Chemistry. Tetrahedron Organic Chemistry Series Volume 19 By Timothy D. W. Claridge (Dyson Perrins Laborotory, Oxford). Pergamon: Oxford. 1999. ix + 384 pp. Hardbound $134.50, ISBN 0-08-042799-5. Paperback $49.50*. *Journal of the American Chemical Society* vol. 122 (2000).

19. Blink, E. J. Meer fysica. *MRI basics* 1–75 (2004).

20. Hazra, A., Lube, G. & Raumer, H. G. Numerical simulation of Bloch equations for dynamic magnetic resonance imaging. *Appl. Numer. Math.* (2018) doi:10.1016/j.apnum.2017.09.007.

21. Theoretical introduction to Nuclear Magnetic Resonance. https://dnangelica.com/dnangelica/index.php/2018/08/14/what-about-nuclear-magnetic-resonance/.

22. Louis, E. *et al.* Detection of lung cancer through metabolic changes measured in blood plasma. *J. Thorac. Oncol.* **11**, 516–523 (2016).

23. Louis, E. *et al.* Phenotyping human blood plasma by 1H-NMR: a robust protocol based on metabolite spiking and its evaluation in breast cancer. *Metabolomics* **11**, 225–236 (2014).

24. Cigarette smoking behaviour and blood metabolomics. https://www.ncbi.nlm.nih.gov/pmc/articles/PMC5100605/.

25. Goldstraw, P. *et al.* The IASLC lung cancer staging project: Proposals for revision of the TNM stage groupings in the forthcoming (eighth) edition of the TNM Classification for lung cancer. *J. Thorac. Oncol.* **11**, 39–51 (2016).

26. Louis, E. *et al.* Metabolic phenotyping of human blood plasma: A powerful tool to discriminate between cancer types? *Ann. Oncol.* **27**, 178–184 (2016).

27. Cross, A. J. *et al.* Metabolites of tobacco smoking and colorectal cancer risk. *Carcinogenesis* (2014) doi:10.1093/carcin/bgu071.

28. Hsu, P. C. *et al.* Feasibility of identifying the tobacco-related global metabolome in blood by UPLC-QTOF-MS. *J. Proteome Res.* **12**, 679–691 (2013).

29. Müller, D. C. *et al.* Metabolomics using GC-TOF-MS followed by subsequent GC-FID and HILIC-MS/MS analysis revealed significantly altered fatty acid and phospholipid species profiles in plasma of smokers. *J. Chromatogr. B Anal. Technol. Biomed. Life Sci.* **966**, 117–126 (2014).

30. Xu, T. *et al.* Effects of smoking and smoking cessation on human serum metabolite profile: Results from the KORA cohort study. *BMC Med.* **11**, (2013).

**Supporting Information**

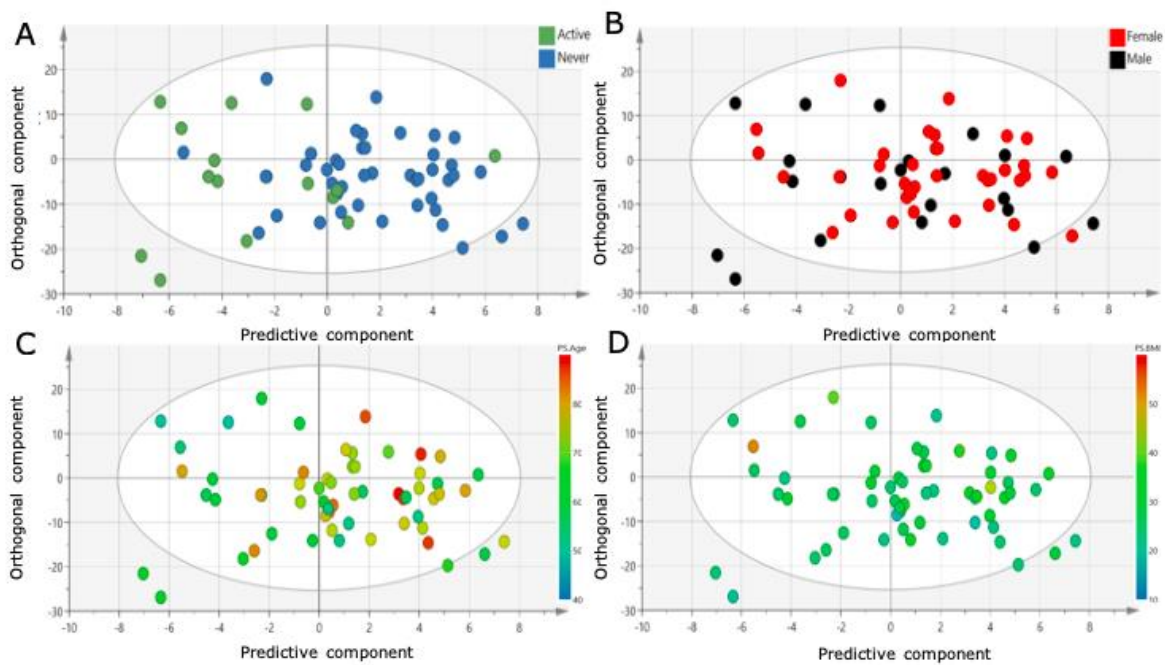

SI Figure 1: OPLS-DA score plot for smoking behavior on the validation data (A), wherein different clinical parameters (B: gender, C: age, D: BMI) are evaluated as possible confounders.

SI Table 1: List of metabolites upon which the classification model was built.

| Pathways | Found metabolites from 10 most contributing variables | Assigned metabolites to build the model |
|---|---|---|
| *Alanine, aspartate and glutamate metabolism* | Ala, Gln, Glu | Ala, Asn, Asp, Citrate, Gln, Glu |
| *Cysteine and methionine metabolism* | Ala, Met, Ser | Ala, Asp, Met, Pyruvate, Ser |
| *Valine, Leucine and Isoleucine degradation* | Acetoacetate, Ile, Leu | Acetoacetate, Ile, Leu, Val |
| *Valine, Leucine and Isoleucine biosynthesis* | Ile, Leu, Thr | Ile, Leu, Pyruvate, Thr, Val |
| *2-oxocarboxylic acid metabolism* | Glu, Leu, Lys, Met, Phe, Tyr | - |
| *Cyanoamino acid metabolism* | Ala, Glu, Ile, Phe, Tyr, Asp, Ser | - |

The 16 variables, which originated from the 10 most contributing variables, made it possible to identify the 6 pathways listed in the first column. The specific metabolites which were present are listed in the second column. In the third column, all the metabolites involved in the pathway (whose integration regions are known by previously performed spiking experiments) are listed. Abbreviations: Ala, alanine; Gln, glutamate; Glu, glutamine; Asn, asparagine; Asp, aspartate; Met, methionine; Ser, serine; Ile, isoleucine; Leu, leucine; Val, valine; Thr, threonine; Phe, phenylalanine; Tyr, tyrosine.